# Multi-Erasure Locally Recoverable Codes Over Small Fields

**Pengfei Huang**[*], **Eitan Yaakobi**[†], and **Paul H. Siegel**[*]

[*]Electrical and Computer Engineering Dept., University of California, San Diego, La Jolla, CA 92093 U.S.A

[†]Computer Science Dept., Technion – Israel Institute of Technology, Haifa 32000, Israel

{*pehuang,psiegel*}@ucsd.edu, *yaakobi*@cs.technion.ac.il

*Abstract*—Erasure codes play an important role in storage systems to prevent data loss. In this work, we study a class of erasure codes called Multi-Erasure Locally Recoverable Codes (ME-LRCs) for storage arrays. Compared to previous related works, we focus on the construction of ME-LRCs over small fields. We first develop upper and lower bounds on the minimum distance of ME-LRCs. Our main contribution is to propose a general construction of ME-LRCs based on generalized tensor product codes, and study their erasure-correcting properties. A decoding algorithm tailored for erasure recovery is given, and correctable erasure patterns are identified. We then prove that our construction yields optimal ME-LRCs with a wide range of code parameters, and present some explicit ME-LRCs over small fields. Finally, we show that generalized integrated interleaving (GII) codes can be treated as a subclass of generalized tensor product codes, thus defining the exact relation between these codes.

## I. INTRODUCTION

Recently, erasure codes with both local and global erasure-correcting properties have received considerable attention [3], [9], [17]–[19], [21], thanks to their promising application in storage systems. The idea behind them is that when only a few erasures occur, these erasures can be corrected fast using only local parities. If the number of erasures exceeds the local erasure-correcting capability, then the global parities are invoked.

In this paper, we consider this kind of erasure codes with both local and global erasure-correcting capabilities for a $\rho \times n_0$ storage array [3], where each row contains some local parities, and additional global parities are distributed in the array. The array structure is suitable for many storage applications. For example, consider a redundant array of independent disks (RAID) type of architecture for solid-state drives (SSDs) [3], [8]. In this scenario, a $\rho \times n_0$ storage array can represent a total of $\rho$ SSDs, each of which contains $n_0$ flash memory chips. Within each SSD, an erasure code is applied to these $n_0$ chips for local protection. In addition, erasure coding is also done across all the SSDs for global protection of all the chips. More specifically, let us give the formal definition of this class of erasure codes as follows.

**Definition 1.** *Consider a code $\mathcal{C}$ over a finite field $\mathbb{F}_q$ consisting of $\rho \times n_0$ arrays such that:*

1) *Each row in each array in $\mathcal{C}$ belongs to a linear local code $\mathcal{C}_0$ with length $n_0$ and minimum distance $d_0$ over $\mathbb{F}_q$.*
2) *Reading the symbols of $\mathcal{C}$ row-wise, $\mathcal{C}$ is a linear code with length $\rho n_0$, dimension $k$, and minimum distance $d$ over $\mathbb{F}_q$.*

*Then, we say that $\mathcal{C}$ is a $(\rho, n_0, k; d_0, d)_q$ Multi-Erasure Locally Recoverable Code (**ME-LRC**).* □

Thus, a $(\rho, n_0, k; d_0, d)_q$ ME-LRC can locally correct $d_0 - 1$ erasures in each row, and is guaranteed to correct a total of $d - 1$ erasures anywhere in the array.

Our work is motivated by a recent work by Blaum and Hetzler [3]. In their work, the authors studied ME-LRCs where each row is a maximum distance separable (MDS) code, and gave code constructions with field size $q \geqslant \max\{\rho, n_0\}$ using generalized integrated interleaving (GII) codes [11], [22], [24]. Our Definition 1 generalizes the definition of the codes in [3] by not requiring each row to be an MDS code. There exist other related works. The ME-LRCs in Definition 1 can be seen as $(r, \delta)$ LRCs with disjoint repair sets. A code $\mathcal{C}$ is called an $(r, \delta)$ LRC [19], if for every coordinate, there exists a punctured code (i.e., a repair set) of $\mathcal{C}$ with support containing this coordinate, whose length is at most $r + \delta - 1$, and whose minimum distance is at least $\delta$. Although the existing constructions [19], [21] for $(r, \delta)$ LRCs with disjoint repair sets can generate ME-LRCs as in Definition 1, they use MDS codes as local codes and require a field size that is at least as large as the code length. A recent work [1] gives explicit constructions of $(r, \delta)$ LRCs with disjoint repair sets over field $\mathbb{F}_q$ from algebraic curves, whose repair sets have size $r + \delta - 1 = \sqrt{q}$ or $r + \delta - 1 = \sqrt{q} + 1$. Partial MDS (PMDS) codes [2] are also related to but different from ME-LRCs in Definition 1. In general, an ME-LRC is not a PMDS code which needs to satisfy more strict requirements. A $\rho \times n_0$ array code is called an $(r; s)$ PMDS code if each row is an $[n_0, n_0 - r, r + 1]_q$ MDS code and whenever any $r$ locations in each row are punctured, the resulting code is also an MDS code with minimum distance $s + 1$. The construction of $(r, s)$ PMDS codes for all $r$ and $s$ with field size $O(n_0^{\rho n_0})$ was known [6]. More recently, a family of PMDS codes with field size $O(\max\{\rho, n_0^{r+s}\}^s)$ was constructed [7].

To the best of our knowledge, however, the construction of *optimal* ME-LRCs over any small field (e.g., the field size less than the length of the local code, or even the binary field) has not been fully explored and solved. The goal of this paper is to study ME-LRCs over small fields. We propose a general construction based on generalized tensor product codes [15], [23], which were first utilized in [12] to construct binary single-erasure LRCs [9], [10], [13], [14], [18], [21]. The contributions of this paper are:

**1)** We extend our previous construction in [12] to the scenario of multi-erasure LRCs over any field. As a result, the construction in [12] can be seen as a special case of the construction proposed in this paper.

2) In contrast to [3], our construction does not require field size $q \geqslant \max\{\rho, n_0\}$, and it can even generate binary ME-LRCs. We derive upper and lower bounds on the minimum distance of ME-LRCs. For $2d_0 \geqslant d$, we show that our construction can produce optimal ME-LRCs with respect to (w.r.t.) the new upper bound on the minimum distance.

3) We present an erasure decoding algorithm and its corresponding correctable erasure patterns, which include the pattern of any $d-1$ erasures. We show that the ME-LRCs from our construction based on Reed-Solomon (RS) codes are optimal w.r.t. certain correctable erasure patterns.

4) So far the *exact* relation between GII codes [3], [22], [24] and generalized tensor product codes has not been fully investigated. We prove that GII codes are a subclass of generalized tensor product codes. As a result, the parameters of a GII code can be obtained by using the known properties of generalized tensor product codes.

The remainder of the paper is organized as follows. In Section II, we study field size dependent upper and lower bounds for ME-LRCs. In Section III, we propose a general construction of ME-LRCs. The erasure-correcting properties of these codes are studied and an erasure decoding algorithm is presented. In Section IV, we study optimal code construction and give several explicit optimal ME-LRCs over different fields. In Section V, we prove that GII codes are a subclass of generalized tensor product codes. Section VI concludes the paper.

Throughout the paper, we use the notation $[n]$ to denote the set $\{1, \ldots, n\}$. For a length-$n$ vector $v$ over $\mathbb{F}_q$ and a set $\mathcal{I} \subseteq [n]$, the vector $v_{\mathcal{I}}$ denotes the restriction of the vector $v$ to coordinates in the set $\mathcal{I}$, and $w_q(v)$ represents the Hamming weight of the vector $v$ over $\mathbb{F}_q$. The transpose of a matrix $H$ is written as $H^T$. For a set $\mathcal{S}$, $|\mathcal{S}|$ represents the cardinality of the set. A linear code $\mathcal{C}$ over $\mathbb{F}_q$ of length $n$, dimension $k$, and minimum distance $d$ will be denoted by $\mathcal{C} = [n, k, d]_q$ or $[n, k, d]_q$ for simplicity. For a code with only one codeword, the minimum distance is defined as $\infty$.

## II. UPPER AND LOWER BOUNDS FOR ME-LRCS

In this section, we derive field size dependent upper and lower bounds on the minimum distance of ME-LRCs. The upper bound obtained here will be used to prove the optimality of our construction for ME-LRCs in the following sections.

Now, we give an upper bound on the minimum distance of a $(\rho, n_0, k; d_0, d)_q$ ME-LRC, by extending the shortening bound for LRCs in [5].

Let $d_{opt}^{(q)}[n, k]$ denote the largest possible minimum distance of a linear code of length $n$ and dimension $k$ over $\mathbb{F}_q$, and let $k_{opt}^{(q)}[n, d]$ denote the largest possible dimension of a linear code of length $n$ and minimum distance $d$ over $\mathbb{F}_q$.

**Lemma 2.** *For any $(\rho, n_0, k; d_0, d)_q$ ME-LRC $\mathcal{C}$, the minimum distance $d$ satisfies*

$$d \leqslant \min_{0 \leqslant x \leqslant \lceil \frac{k}{k^*} \rceil - 1, \, x \in \mathbb{Z}} \left\{ d_{opt}^{(q)}[\rho n_0 - x n_0, k - x k^*] \right\}, \quad (1)$$

*and the dimension satisfies*

$$k \leqslant \min_{0 \leqslant x \leqslant \lceil \frac{k}{k^*} \rceil - 1, \, x \in \mathbb{Z}} \left\{ x k^* + k_{opt}^{(q)}[\rho n_0 - x n_0, d] \right\}, \quad (2)$$

*where $k^* = k_{opt}^{(q)}[n_0, d_0]$.*

*Proof:* See Appendix A. ∎

An asymptotic lower bound for ME-LRCs with local MDS codes was given in [1]. Here, by simply adapting the Gilbert-Varshamov (GV) bound [20], we have the following GV-like lower bound on ME-LRCs of finite length without specifying local codes.

**Lemma 3.** *A $(\rho, n_0, k; \geqslant d_0, \geqslant d)_q$ ME-LRC $\mathcal{C}$ exists, if*

$$\sum_{i=0}^{d-2} \binom{\rho(n_0 - \lceil \log_q \left( \sum_{j=0}^{d_0-2} \binom{n_0-1}{j}(q-1)^j \right) \rceil) - 1}{i}(q-1)^i$$
$$< q^{\rho(n_0 - \lceil \log_q \left( \sum_{j=0}^{d_0-2} \binom{n_0-1}{j}(q-1)^j \right) \rceil) - k}. \quad (3)$$

*Proof:* See Appendix B. ∎

## III. ME-LRCS FROM GENERALIZED TENSOR PRODUCT CODES: CONSTRUCTION AND DECODING

Tensor product codes, first proposed by Wolf in [23], are a family of binary error-correcting codes defined by a parity-check matrix that is the tensor product of the parity-check matrices of two constituent codes. Later, they were generalized in [15]. In this section, we first introduce generalized tensor product codes over $\mathbb{F}_q$. Then, we give a general construction of ME-LRCs from generalized tensor product codes. The minimum distance of the constructed ME-LRCs is determined, a decoding algorithm tailored for erasure correction is proposed, and corresponding correctable erasure patterns are studied.

### A. Generalized Tensor Product Codes over $\mathbb{F}_q$

We start by presenting the tensor product operation of two matrices $H'$ and $H''$. Let $H'$ be the parity-check matrix of a code with length $n'$ and dimension $n' - v$ over $\mathbb{F}_q$. The matrix $H'$ can be considered as a $v$ (row) by $n'$ (column) matrix over $\mathbb{F}_q$ or as a 1 (row) by $n'$ (column) matrix of elements from $\mathbb{F}_{q^v}$. Let $H'$ be the vector $H' = [h'_1 \; h'_2 \; \cdots \; h'_{n'}]$, where $h'_j, 1 \leqslant j \leqslant n'$, are elements of $\mathbb{F}_{q^v}$. Let $H''$ be the parity-check matrix of a code of length $\ell$ and dimension $\ell - \lambda$ over $\mathbb{F}_{q^v}$. We denote $H''$ by

$$H'' = \begin{bmatrix} h''_{11} & \cdots & h''_{1\ell} \\ \vdots & \ddots & \vdots \\ h''_{\lambda 1} & \cdots & h''_{\lambda \ell} \end{bmatrix},$$

where $h''_{ij}, 1 \leqslant i \leqslant \lambda$ and $1 \leqslant j \leqslant \ell$, are elements of $\mathbb{F}_{q^v}$.

The tensor product of the matrices $H''$ and $H'$ is defined as

$$H_{TP} = H'' \bigotimes H' = \begin{bmatrix} h''_{11} H' & \cdots & h''_{1\ell} H' \\ \vdots & \ddots & \vdots \\ h''_{\lambda 1} H' & \cdots & h''_{\lambda \ell} H' \end{bmatrix},$$

where $h''_{ij}H' = [h''_{ij}h'_1 \ h''_{ij}h'_2 \ \cdots \ h''_{ij}h'_{n'}]$, $1 \leq i \leq \lambda$ and $1 \leq j \leq \ell$, and the products of elements are calculated according to the rules of multiplication for elements over $\mathbb{F}_{q^v}$. The matrix $H_{TP}$ will be considered as a $v\lambda \times n'\ell$ matrix of elements from $\mathbb{F}_q$, thus defining a tensor product code over $\mathbb{F}_q$. We provide an example to illustrate these operations.

**Example 1.** (cf. [23]) Let $H''$ be the following parity-check matrix over $\mathbb{F}_4$ for a $[5,3,3]_4$ code where $\alpha$ is a primitive element of $\mathbb{F}_4$,

$$H'' = \begin{bmatrix} \alpha^0 & 0 & \alpha^0 & \alpha^0 & \alpha^0 \\ 0 & \alpha^0 & \alpha^0 & \alpha^1 & \alpha^2 \end{bmatrix}.$$

Let $H'$ be the following parity-check matrix over $\mathbb{F}_2$ for a $[3,1,3]_2$ Hamming code,

$$H' = \begin{bmatrix} 1 & 0 & 1 \\ 0 & 1 & 1 \end{bmatrix}.$$

Representing the elements of $\mathbb{F}_4$ as $\alpha^0 = \begin{bmatrix} 1 \\ 0 \end{bmatrix}$, $\alpha^1 = \begin{bmatrix} 0 \\ 1 \end{bmatrix}$, $\alpha^2 = \begin{bmatrix} 1 \\ 1 \end{bmatrix}$, and $0 = \begin{bmatrix} 0 \\ 0 \end{bmatrix}$, we have

$$H_{TP} = H'' \bigotimes H'$$
$$= \begin{bmatrix} \alpha^0 \alpha^1 \alpha^2 & 0 \ 0 \ 0 & \alpha^0 \alpha^1 \alpha^2 & \alpha^0 \alpha^1 \alpha^2 & \alpha^0 \alpha^1 \alpha^2 \\ 0 \ 0 \ 0 & \alpha^0 \alpha^1 \alpha^2 & \alpha^0 \alpha^1 \alpha^2 & \alpha^1 \alpha^2 \alpha^0 & \alpha^2 \alpha^0 \alpha^1 \end{bmatrix}.$$

Using the same symbol-to-binary vector mapping, we represent $H_{TP}$ over $\mathbb{F}_2$ as

$$H_{TP} = \begin{bmatrix} 1 \ 0 \ 1 & 0 \ 0 \ 0 & 1 \ 0 \ 1 & 1 \ 0 \ 1 & 1 \ 0 \ 1 \\ 0 \ 1 \ 1 & 0 \ 0 \ 0 & 0 \ 1 \ 1 & 0 \ 1 \ 1 & 0 \ 1 \ 1 \\ 0 \ 0 \ 0 & 1 \ 0 \ 1 & 1 \ 0 \ 1 & 0 \ 1 \ 1 & 1 \ 1 \ 0 \\ 0 \ 0 \ 0 & 0 \ 1 \ 1 & 0 \ 1 \ 1 & 1 \ 1 \ 0 & 1 \ 0 \ 1 \end{bmatrix},$$

which defines a binary $[15, 11, 3]_2$ code. □

Our construction of ME-LRCs is based on generalized tensor product codes [15]. Define the matrices $H'_i$ and $H''_i$ for $i = 1, 2, \ldots, \mu$ as follows. The matrix $H'_i$ is a $v_i \times n'$ matrix over $\mathbb{F}_q$ such that the $(v_1 + v_2 + \cdots + v_i) \times n'$ matrix

$$B_i = \begin{bmatrix} H'_1 \\ H'_2 \\ \vdots \\ H'_i \end{bmatrix}$$

is a parity-check matrix of an $[n', n' - v_1 - v_2 - \cdots - v_i, d'_i]_q$ code $\mathcal{C}'_i$, where $d'_1 \leq d'_2 \leq \cdots \leq d'_i$. The matrix $H''_i$ is a $\lambda_i \times \ell$ matrix over $\mathbb{F}_{q^{v_i}}$, which is a parity-check matrix of an $[\ell, \ell - \lambda_i, \delta_i]_{q^{v_i}}$ code $\mathcal{C}''_i$.

We define a $\mu$-level generalized tensor product code over $\mathbb{F}_q$ as a linear code having a parity-check matrix over $\mathbb{F}_q$ in the form of the following $\mu$-level tensor product structure

$$H = \begin{bmatrix} H''_1 \otimes H'_1 \\ H''_2 \otimes H'_2 \\ \vdots \\ H''_\mu \otimes H'_\mu \end{bmatrix}. \quad (4)$$

As the matrix $H_{TP}$, each level in the matrix $H$ is obtained by operations over $\mathbb{F}_q$ and its extension field. We denote this code by $\mathcal{C}^\mu_{GTP}$. Its length is $n_t = n'\ell$ and the dimension is $k_t = n_t - \sum_{i=1}^{\mu} v_i \lambda_i$.

By adapting Theorem 2 in [15] from the field $\mathbb{F}_2$ to $\mathbb{F}_q$, we directly have the following theorem on the minimum distance of $\mathcal{C}^\mu_{GTP}$ over $\mathbb{F}_q$.

**Theorem 4.** *The minimum distance $d_t$ of a generalized tensor product code $\mathcal{C}^\mu_{GTP}$ over $\mathbb{F}_q$ satisfies*

$$d_t \geq \min\{\delta_1, \delta_2 d'_1, \delta_3 d'_2, \ldots, \delta_\mu d'_{\mu-1}, d'_\mu\}.$$

*Proof:* See Appendix C. ∎

### B. Construction of ME-LRCs

Now, we present a general construction of ME-LRCs based on generalized tensor product codes.

**Construction A**

**Step 1:** Choose $v_i \times n'$ matrices $H'_i$ over $\mathbb{F}_q$ and $\lambda_i \times \ell$ matrices $H''_i$ over $\mathbb{F}_{q^{v_i}}$, for $i = 1, 2, \ldots, \mu$, which satisfy the following two properties:
1) The parity-check matrix $H''_1 = \mathbf{I}_{\ell \times \ell}$, i.e., an $\ell \times \ell$ identity matrix.
2) The matrices $H'_i$ (or $B_i$), $1 \leq i \leq \mu$, and $H''_j$, $2 \leq j \leq \mu$, are chosen such that $d'_\mu \leq \delta_j d'_{j-1}$, for $j = 2, 3, \ldots, \mu$.

**Step 2:** Generate a parity-check matrix $H$ over $\mathbb{F}_q$ according to (4) with the matrices $H'_i$ and $H''_i$, for $i = 1, 2, \ldots, \mu$. The constructed code corresponding to the parity-check matrix $H$ is referred to as $\mathcal{C}_A$. ∎

**Theorem 5.** *The code $\mathcal{C}_A$ is a $(\rho, n_0, k; d_0, d)_q$ ME-LRC with parameters $\rho = \ell$, $n_0 = n'$, $k = n'\ell - \sum_{i=1}^{\mu} v_i \lambda_i$, $d_0 = d'_1$, and $d = d'_\mu$.*

*Proof:* According to Construction A, the code parameters $\rho$, $n_0$, $k$, and $d_0$ can be easily determined. In the following, we prove that $d = d'_\mu$.

Since $\delta_1 = \infty$ ($H''_1$ is the identity matrix) and $d'_\mu \leq \delta_i d'_{i-1}$ for all $i = 2, 3, \ldots, \mu$, from Theorem 4, $d \geq d'_\mu$.

Now, we show that $d \leq d'_\mu$. For $i = 1, 2, \ldots, \mu$, let $H'_i = [h'_1(i), \ldots, h'_{n'}(i)]$ over $\mathbb{F}_{q^{v_i}}$, and let $[h''_{11}(i), \ldots, h''_{\lambda_i 1}(i)]^T$ over $\mathbb{F}_{q^{v_i}}$ be the first column of $H''_i$. Since the code with parity-check matrix $B_\mu$ has minimum distance $d'_\mu$, there exist $d'_\mu$ columns of $B_\mu$, say in the set of positions $J = \{b_1, b_2, \ldots, b_{d'_\mu}\}$, which are linearly dependent; that is, $\sum_{j \in J} \alpha_j h'_j(i) = 0$, for some $\alpha_j \in \mathbb{F}_q$, for all $i = 1, 2, \ldots, \mu$. Thus, we have $\sum_{j \in J} \alpha_j h''_{p1}(i) h'_j(i) = h''_{p1}(i) \left( \sum_{j \in J} \alpha_j h'_j(i) \right) = 0$, for $p = 1, 2, \ldots, \lambda_i$ and $i = 1, 2, \ldots, \mu$. That is, the columns in positions $b_1, b_2, \ldots, b_{d'_\mu}$ of $H$ are linearly dependent. ∎

### C. Erasure Decoding and Correctable Erasure Patterns

We present a decoding algorithm for the ME-LRC $\mathcal{C}_A$ from Construction A, tailored for erasure correction. The decoding algorithm for error correction for generalized tensor product codes can be found in [15].

Let the symbol ? represent an erasure and "e" denote a decoding failure. The erasure decoder $\mathcal{D}_A : (\mathbb{F}_q \cup \{?\})^{n'\ell} \to$

$\mathcal{C}_A \cup \{\text{"e"}\}$ for an ME-LRC $\mathcal{C}_A$ consists of two kinds of component decoders $\mathcal{D}'_i$ and $\mathcal{D}''_i$ for $i = 1, 2, \ldots, \mu$ described below.

**a)** First, the decoder for a coset of the code $\mathcal{C}'_i$ with parity-check matrix $B_i$, $i = 1, 2, \ldots, \mu$, is denoted by

$$\mathcal{D}'_i : (\mathbb{F}_q \cup \{?\})^{n'} \times (\mathbb{F}_q \cup \{?\})^{\sum_{j=1}^{i} v_j} \to (\mathbb{F}_q \cup \{?\})^{n'}$$

which uses the following decoding rule: for a length-$n'$ input vector $\bm{y}'$, and a length-$\sum_{j=1}^{i} v_j$ syndrome vector $\bm{s}'$ without erasures, if $\bm{y}'$ agrees with exactly one codeword $\bm{c}' \in \mathcal{C}'_i + \bm{e}$ on the entries with values in $\mathbb{F}_q$, where the vector $\bm{e}$ is a coset leader determined by both the code $\mathcal{C}'_i$ and the syndrome vector $\bm{s}'$, i.e., $\bm{s}' = \bm{e} B_i^T$, then $\mathcal{D}'_i(\bm{y}', \bm{s}') = \bm{c}'$; otherwise, $\mathcal{D}'_i(\bm{y}', \bm{s}') = \bm{y}'$. Therefore, if the length-$n'$ input vector $\bm{y}'$ is a codeword in $\mathcal{C}'_i + \bm{e}$ with $d'_i - 1$ or less erasures and the syndrome vector $\bm{s}'$ is not erased, then the decoder $\mathcal{D}'_i$ can return the correct codeword.

**b)** Second, the decoder for the code $\mathcal{C}''_i$ with parity-check matrix $H''_i$, $i = 1, 2, \ldots, \mu$, is denoted by

$$\mathcal{D}''_i : (\mathbb{F}_{q^{v_i}} \cup \{?\})^{\ell} \to (\mathbb{F}_{q^{v_i}} \cup \{?\})^{\ell}$$

which uses the following decoding rule: for a length-$\ell$ input vector $\bm{y}''$, if $\bm{y}''$ agrees with exactly one codeword $\bm{c}'' \in \mathcal{C}''_i$ on the entries with values in $\mathbb{F}_{q^{v_i}}$, then $\mathcal{D}''_i(\bm{y}'') = \bm{c}''$; otherwise, $\mathcal{D}''_i(\bm{y}'') = \bm{y}''$. Therefore, if the length-$\ell$ input vector $\bm{y}''$ is a codeword in $\mathcal{C}''_i$ with $\delta_i - 1$ or less erasures, then the decoder $\mathcal{D}''_i$ can successfully return the correct codeword.

The erasure decoder $\mathcal{D}_A$ for the code $\mathcal{C}_A$ is summarized in Algorithm 1 below. Let the input word of length $n'\ell$ for the decoder $\mathcal{D}_A$ be $\bm{y} = (\bm{y}_1, \bm{y}_2, \ldots, \bm{y}_\ell)$, where each component $\bm{y}_i \in (\mathbb{F}_q \cup \{?\})^{n'}$, $i = 1, \ldots, \ell$. The vector $\bm{y}$ is an erased version of a codeword $\bm{c} = (\bm{c}_1, \bm{c}_2, \ldots, \bm{c}_\ell) \in \mathcal{C}_A$.

---

**Algorithm 1: Decoding Procedure of Decoder $\mathcal{D}_A$**

**Input:** received word $\bm{y} = (\bm{y}_1, \bm{y}_2, \ldots, \bm{y}_\ell)$.
**Output:** codeword $\bm{c} \in \mathcal{C}_A$ or a decoding failure "e".

1. Let $\bm{s}^1_j = \bm{0}$, for $j = 1, 2, \ldots, \ell$.
2. $\hat{\bm{c}} = (\hat{\bm{c}}_1, \ldots, \hat{\bm{c}}_\ell) = \left( \mathcal{D}'_1(\bm{y}_1, \bm{s}^1_1), \ldots, \mathcal{D}'_1(\bm{y}_\ell, \bm{s}^1_\ell) \right)$.
3. Let $\mathcal{F} = \{j \in [\ell] : \hat{\bm{c}}_j \text{ contains ?}\}$.
4. **For** $i = 2, \ldots, \mu$
   - If $\mathcal{F} \neq \emptyset$, do the following steps; otherwise go to step 5.
   - $(\bm{s}^i_1, \ldots, \bm{s}^i_\ell) = \mathcal{D}''_i\left( \hat{\bm{c}}_1 H'^T_i, \ldots, \hat{\bm{c}}_\ell H'^T_i \right)$.
   - $\hat{\bm{c}}_j = \mathcal{D}'_i\left( \hat{\bm{c}}_j, (\bm{s}^1_j, \ldots, \bm{s}^i_j) \right)$ for $j \in \mathcal{F}$; $\hat{\bm{c}}_j$ remains the same for $j \in [\ell] \setminus \mathcal{F}$.
   - Update $\mathcal{F} = \{j \in [\ell] : \hat{\bm{c}}_j \text{ contains ?}\}$.
   **end**
5. If $\mathcal{F} = \emptyset$, let $\bm{c} = \hat{\bm{c}}$ and output $\bm{c}$; otherwise return "e".

---

In Algorithm 1, we use the following rules for operations which involve the symbol ?: 1) Addition $+$: for any element $\gamma \in \mathbb{F}_q \cup \{?\}$, $\gamma + ? = ?$. 2) Multiplication $\times$: for any element $\gamma \in \mathbb{F}_q \cup \{?\} \setminus \{0\}$, $\gamma \times ? = ?$, and $0 \times ? = 0$. 3) If a length-$n$ vector $\bm{x}$, $\bm{x} \in (\mathbb{F}_q \cup \{?\})^n$, contains an entry ?, then $\bm{x}$ is considered as the symbol ? in the set $\mathbb{F}_{q^n} \cup \{?\}$. Similarly, the symbol ? in the set $\mathbb{F}_{q^n} \cup \{?\}$ is treated as a length-$n$ vector whose entries are all ?.

To describe correctable erasure patterns, we use the following notation. Let $w_e(\bm{v})$ denote the number of erasures ? in the vector $\bm{v}$. For a received word $\bm{y} = (\bm{y}_1, \bm{y}_2, \ldots, \bm{y}_\ell)$, let $N_\tau = |\{\bm{y}_m : w_e(\bm{y}_m) \geq d'_\tau, 1 \leq m \leq \ell\}|$ for $1 \leq \tau \leq \mu$.

**Theorem 6.** *The decoder $\mathcal{D}_A$ for a $(\rho, n_0, k; d_0, d)_q$ ME-LRC $\mathcal{C}_A$ can correct any received word $\bm{y}$ that satisfies the following condition:*

$$N_\tau \leq \delta_{\tau+1} - 1, \; \forall \, 1 \leq \tau \leq \mu, \quad (5)$$

*where $\delta_{\mu+1}$ is defined to be 1.*

*Proof:* See Appendix D. ∎

The following corollary follows from Theorem 6.

**Corollary 7.** *The decoder $\mathcal{D}_A$ for a $(\rho, n_0, k; d_0, d)_q$ ME-LRC $\mathcal{C}_A$ can correct any received word $\bm{y}$ with less than $d$ erasures.*

*Proof:* See Appendix E. ∎

## IV. OPTIMAL CONSTRUCTION AND EXPLICIT ME-LRCS OVER SMALL FIELDS

In this section, we study the optimality of Construction A, and also present several explicit ME-LRCs that are optimal over different fields.

### A. Optimal Construction

We show how to construct ME-LRCs which are optimal w.r.t. the bound (1) by adding more constraints to Construction A. To this end, we specify the choice of the matrices in Construction A. This specification, referred to as **Design I**, is as follows.

1) $H'_1$ is the parity-check matrix of an $[n', n' - v_1, d'_1]_q$ code which satisfies $k^{(q)}_{opt}[n', d'_1] = n' - v_1$.
2) $B_\mu$ is the parity-check matrix of an $[n', n' - \sum_{i=1}^{\mu} v_i, d'_\mu]_q$ code with $d^{(q)}_{opt}[n', n' - \sum_{i=1}^{\mu} v_i] = d'_\mu$.
3) The minimum distances satisfy $d'_\mu \leq 2 d'_1$.
4) $H''_i$ is an all-one vector of length $\ell$ over $\mathbb{F}_{q^{v_i}}$, i.e., the parity-check matrix of a parity code with minimum distance $\delta_i = 2$, for all $i = 2, \ldots, \mu$. ∎

**Theorem 8.** *The code $\mathcal{C}_A$ from Construction A with Design I is a $(\rho = \ell, n_0 = n', k = n'\ell - v_1 \ell - \sum_{i=2}^{\mu} v_i; d_0 = d'_1, d = d'_\mu)_q$ ME-LRC, which is optimal with respect to the bound (1).*

*Proof:* From Theorem 5, the code parameters $\rho, n_0, k, d_0$, and $d$ can be determined. We have $k^* = k^{(q)}_{opt}[n', d'_1] = n' - v_1$. Setting $x = \ell - 1$, we get

$$d \leq \min_{0 \leq x \leq \lceil \frac{k}{k^*} \rceil - 1} \left\{ d^{(q)}_{opt}[\rho n_0 - x n_0, k - x k^*] \right\}$$

$$\leq d^{(q)}_{opt}[\ell n' - (\ell-1)n', k - (\ell-1)k^*]$$

$$= d^{(q)}_{opt}\left[n', n' - \sum_{i=1}^{\mu} v_i\right] = d'_\mu.$$

This proves that $\mathcal{C}_A$ achieves the bound (1). ∎

## B. Explicit ME-LRCs from Construction A

Our construction is very flexible and can generate many ME-LRCs over different fields. In the following, we present several examples.

*1) ME-LRCs with local extended BCH codes over $\mathbb{F}_2$*

From the structure of BCH codes [20], there exists a chain of nested binary extended BCH codes: $\mathcal{C}_3 = [2^m, 2^m - 1 - 3m, 8]_2 \subset \mathcal{C}_2 = [2^m, 2^m - 1 - 2m, 6]_2 \subset \mathcal{C}_1 = [2^m, 2^m - 1 - m, 4]_2$.

Let the matrices $B_1$, $B_2$, and $B_3$ be the parity-check matrices of $\mathcal{C}_1$, $\mathcal{C}_2$, and $\mathcal{C}_3$, respectively.

**Example 2.** For $\mu = 3$, in Construction A, we use the above matrices $B_1$, $B_2$, and $B_3$. We also choose $H_2''$ and $H_3''$ to be the all-one vector of length $\ell$ over $\mathbb{F}_{2^m}$.

From Theorem 5, the corresponding $(\rho, n_0, k; d_0, d)_2$ ME-LRC $\mathcal{C}_A$ has parameters $\rho = \ell$, $n_0 = 2^m$, $k = 2^m \ell - (m+1)\ell - 2m$, $d_0 = 4$, and $d = 8$. This code satisfies the requirements of Design I. Thus, from Theorem 8, it is optimal w.r.t. the bound (1). □

*2) ME-LRCs with local algebraic geometry codes over $\mathbb{F}_4$*

We use a class of algebraic geometry codes called Hermitian codes [25] to construct ME-LRCs.

From the construction of Hermitian codes [25], there exists a chain of nested 4-ary Hermitian codes: $\mathcal{C}_H(1) = [8, 1, 8]_4 \subset \mathcal{C}_H(2) = [8, 2, 6]_4 \subset \mathcal{C}_H(3) = [8, 3, 5]_4 \subset \mathcal{C}_H(4) = [8, 4, 4]_4 \subset \mathcal{C}_H(5) = [8, 5, 3]_4 \subset \mathcal{C}_H(6) = [8, 6, 2]_4 \subset \mathcal{C}_H(7) = [8, 7, 2]_4$.

Now, let the matrices $B_1$, $B_2$, $B_3$, and $B_4$ be the parity-check matrices of $\mathcal{C}_H(4)$, $\mathcal{C}_H(3)$, $\mathcal{C}_H(2)$, and $\mathcal{C}_H(1)$, respectively. Let $H_i''$, $i = 2, 3, 4$, be the all-one vector of length $\ell$ over $\mathbb{F}_4$.

**Example 3.** For $\mu = 2$, in Construction A, we use the above matrices $B_1$, $B_2$, and $H_2''$. From Theorem 5, the corresponding $(\rho, n_0, k; d_0, d)_4$ ME-LRC $\mathcal{C}_A$ has parameters $\rho = \ell$, $n_0 = 8$, $k = 4\ell - 1$, $d_0 = 4$, and $d = 5$.

For $\mu = 3$, in Construction A, we use the above matrices $B_1$, $B_2$, $B_3$, $H_2''$, and $H_3''$. From Theorem 5, the corresponding $(\rho, n_0, k; d_0, d)_4$ ME-LRC $\mathcal{C}_A$ has parameters $\rho = \ell$, $n_0 = 8$, $k = 4\ell - 2$, $d_0 = 4$, and $d = 6$.

For $\mu = 4$, in Construction A, we use the above matrices $B_i$, $i = 1, \ldots, 4$, and $H_j''$, $j = 2, 3, 4$. From Theorem 5, the corresponding $(\rho, n_0, k; d_0, d)_4$ ME-LRC $\mathcal{C}_A$ has parameters $\rho = \ell$, $n_0 = 8$, $k = 4\ell - 3$, $d_0 = 4$, and $d = 8$.

All of the above three families of ME-LRCs over $\mathbb{F}_4$ are optimal w.r.t. the bound (1). □

*3) ME-LRCs with local singly-extended Reed-Solomon codes over $\mathbb{F}_q$*

Let $n' \leqslant q$ and $\alpha$ be a primitive element of $\mathbb{F}_q$. We choose $H_1'$ to be the parity-check matrix of an $[n', n' - d_1' + 1, d_1']_q$ singly-extended RS code, namely

$$H_1' = \begin{bmatrix} 1 & 1 & \cdots & 1 & 1 \\ 1 & \alpha & \cdots & \alpha^{n'-2} & 0 \\ \vdots & \vdots & \ddots & \vdots & \vdots \\ 1 & \alpha^{d_1'-2} & \cdots & \alpha^{(n'-2)(d_1'-2)} & 0 \end{bmatrix}.$$

For $i = 2, 3, \ldots, \mu$, we choose $H_i'$ to be

$$H_i' = \begin{bmatrix} 1 & \alpha^{d_{i-1}'-1} & \cdots & \alpha^{(n'-2)(d_{i-1}'-1)} & 0 \\ \vdots & \vdots & \ddots & \vdots & \vdots \\ 1 & \alpha^{d_i'-2} & \cdots & \alpha^{(n'-2)(d_i'-2)} & 0 \end{bmatrix},$$

where $d_1' < d_2' < \cdots < d_\mu'$. We also require that

$$\delta_i = \lceil \frac{d_\mu'}{d_{i-1}'} \rceil = \lceil \frac{d_\mu'}{d_{i-1}'+1} \rceil = \cdots = \lceil \frac{d_\mu'}{d_i'-1} \rceil, \forall i = 2, \ldots, \mu$$

and $\delta_2 > \delta_3 > \cdots > \delta_\mu$.

For $i = 2, 3, \ldots, \mu$, let $H_i''$ be the parity-check matrix of an $[\ell, \ell - \delta_i + 1, \delta_i = \lceil \frac{d_\mu'}{d_{i-1}'} \rceil]_{q^{v_i}}$ MDS code, which exists whenever $\ell \leqslant q^{v_i}$, where $v_i = d_i' - d_{i-1}'$. Note that for an MDS code with minimum distance 2, the code length can be arbitrarily long.

**Example 4.** We use the above chosen matrices $H_i'$ and $H_i''$ for Construction A. The corresponding $(\rho, n_0, k; d_0, d)_q$ ME-LRC $\mathcal{C}_A$ has parameters $\rho = \ell$, $n_0 = n'$, $k = (n' - d_1' + 1)\ell - \sum_{i=2}^{\mu}(\lceil \frac{d_\mu'}{d_{i-1}'} \rceil - 1)(d_i' - d_{i-1}')$, $d_0 = d_1'$, and $d = d_\mu'$; the field size $q$ satisfies $q \geqslant \max\{q', n'\}$, where $q' = \max_{i=2,\ldots,\mu}\{\lceil \ell^{\frac{1}{d_i'-d_{i-1}'}} \rceil\}$.

When $\mu = 2$ and $d_1' < d_2' \leqslant 2d_1'$, the corresponding $(\rho, n_0, k; d_0, d)_q$ ME-LRC $\mathcal{C}_A$ has parameters $\rho = \ell$, $n_0 = n'$, $k = (n' - d_1' + 1)\ell - (d_2' - d_1')$, $d_0 = d_1'$, and $d = d_2'$; the field size $q$ needs to satisfy $q \geqslant n'$. Since $\mathcal{C}_A$ satisfies the requirements of Design I, from Theorem 8, it is optimal w.r.t. the bound (1). □

The following theorem shows that the $\mu$-level ME-LRC $\mathcal{C}_A$ constructed in Example 4 is optimal in the sense of possessing the largest possible dimension among all codes with the same erasure-correcting capability.

**Theorem 9.** Let $\mathcal{C}$ be a code of length $\ell n'$ and dimension $k$ over $\mathbb{F}_q$. Each codeword in $\mathcal{C}$ consists of $\ell$ sub-blocks, each of length $n'$. Assume that $\mathcal{C}$ corrects all erasure patterns satisfying the condition in (5), where $\delta_\tau = \lceil \frac{d_\mu'}{d_{\tau-1}'} \rceil$ for $2 \leqslant \tau \leqslant \mu$. Then, we must have dimension $k \leqslant (n' - d_1' + 1)\ell - \sum_{i=2}^{\mu}(\lceil \frac{d_\mu'}{d_{i-1}'} \rceil - 1)(d_i' - d_{i-1}')$.

*Proof:* The proof is based on contradiction.

Let each codeword in $\mathcal{C}$ correspond to an $\ell \times n'$ array. We index the coordinates of the array row by row from number 1 to $\ell n'$. Let $\mathcal{I}_1$ be the set of coordinates defined by $\mathcal{I}_1 = \{(i-1)n' + j : \delta_2 - 1 < i \leqslant \ell, 1 \leqslant j \leqslant d_1' - 1\}$. For $2 \leqslant \tau \leqslant \mu$, let $\mathcal{I}_\tau$ be the set of coordinates given by $\mathcal{I}_\tau =$

$\{(i-1)n' + j : \delta_{\tau+1} - 1 < i \leqslant \delta_\tau - 1, \ 1 \leqslant j \leqslant d'_\tau - 1\}$, where $\delta_{\mu+1}$ is defined to be 1. Let $\mathcal{I}$ be the set of all the coordinates of the array.

By calculation, we have $|\mathcal{I}\setminus(\mathcal{I}_1 \cup \mathcal{I}_2 \cup \cdots \cup \mathcal{I}_\mu)| = (n' - d'_1 + 1)\ell - \sum_{i=2}^{\mu}(\lceil \frac{d'_\mu}{d'_{i-1}} \rceil - 1)(d'_i - d'_{i-1})$. Now, assume that $k > (n' - d'_1 + 1)\ell - \sum_{i=2}^{\mu}(\lceil \frac{d'_\mu}{d'_{i-1}} \rceil - 1)(d'_i - d'_{i-1})$. Then, there exist at least two distinct codewords $c'$ and $c''$ in $\mathcal{C}$ that agree on the coordinates in the set $\{i : i \in \mathcal{I}\setminus(\mathcal{I}_1 \cup \mathcal{I}_2 \cup \cdots \cup \mathcal{I}_\mu)\}$. We erase the values on the coordinates in the set $\{i : i \in \mathcal{I}_1 \cup \mathcal{I}_2 \cup \cdots \cup \mathcal{I}_\mu\}$ of both $c'$ and $c''$. This erasure pattern satisfies the condition in (5). Since $c'$ and $c''$ are distinct, this erasure pattern is uncorrectable. Thus, our assumption that $k > (n' - d'_1 + 1)\ell - \sum_{i=2}^{\mu}(\lceil \frac{d'_\mu}{d'_{i-1}} \rceil - 1)(d'_i - d'_{i-1})$ is violated. ∎

**Remark 1.** The construction by Blaum and Hetzler [3] based on GII codes cannot generate ME-LRCs constructed in Examples 2 and 3. For the ME-LRC in Example 4, since the local code is the singly-extended RS code, the construction in [3] can also be used to produce an ME-LRC that has the same code parameters $\rho$, $n_0$, $k$, $d_0$ and $d$ as those of the ME-LRC $\mathcal{C}_A$ from our construction. However, the construction in [3] requires the field size $q$ to satisfy $q \geqslant \max\{\ell, n'\}$, which in general is larger than that in our construction.

## V. RELATION TO GENERALIZED INTEGRATED INTERLEAVING CODES

Integrated interleaving (II) codes were first introduced in [11] as a two-level error-correcting scheme for data storage applications, and were then extended in [22] and more recently in [24] as generalized integrated interleaving (GII) codes for multi-level data protection.

The main difference between GII codes and generalized tensor product codes is that a generalized tensor product code over $\mathbb{F}_q$ is defined by operations over the base field $\mathbb{F}_q$ and also its extension field, as shown in (4); in contrast, a GII code over $\mathbb{F}_q$ is defined over the same field $\mathbb{F}_q$. As a result, generalized tensor product codes are more flexible than GII codes, and generally GII codes cannot be used to construct ME-LRCs over very small fields, e.g., the binary field.

The goal of this section is to study the exact relation between generalized tensor product codes and GII codes. We will show that GII codes are in fact a subclass of generalized tensor product codes. The idea is to reformulate the parity-check matrix of a GII code into the form of a parity-check matrix of a generalized tensor product code. Establishing this relation allows some code properties of GII codes to be obtained directly from known results about generalized tensor product codes. We start by considering the II codes, a two-level case of GII codes, to illustrate our idea.

### A. Integrated Interleaving Codes

We follow the definition of II codes in [11]. Let $\mathcal{C}_i$, $i = 1, 2$, be $[n, k_i, d_i]_q$ linear codes over $\mathbb{F}_q$ such that $\mathcal{C}_2 \subset \mathcal{C}_1$ and $d_2 > d_1$. An II code $\mathcal{C}_{II}$ is defined as follows:

$$\mathcal{C}_{II} = \Big\{ c = (c_0, c_1, \ldots, c_{m-1}) : c_i \in \mathcal{C}_1, 0 \leqslant i < m, \text{ and } \sum_{i=0}^{m-1} \alpha^{bi} c_i \in \mathcal{C}_2, \ b = 0, 1, \ldots, \gamma - 1 \Big\}, \quad (6)$$

where $\alpha$ is a primitive element of $\mathbb{F}_q$ and $\gamma < m \leqslant q - 1$.

According to the above definition, it is known that the parity-check matrix of $\mathcal{C}_{II}$ is

$$H_{II} = \begin{bmatrix} I & \otimes & H_1 \\ \Gamma_2 & \otimes & H_2 \end{bmatrix}, \quad (7)$$

where $\otimes$ denotes the Kronecker product. The matrices $H_1$ and $\begin{bmatrix} H_1 \\ H_2 \end{bmatrix}$ over $\mathbb{F}_q$ are the parity-check matrices of $\mathcal{C}_1$ and $\mathcal{C}_2$, respectively, the matrix $I$ over $\mathbb{F}_q$ is an $m \times m$ identity matrix, and $\Gamma_2$ over $\mathbb{F}_q$ is the parity-check matrix of an $[m, m - \gamma, \gamma + 1]_q$ code in the following form

$$\Gamma_2 = \begin{bmatrix} 1 & 1 & \cdots & 1 \\ 1 & \alpha & \cdots & \alpha^{m-1} \\ 1 & \alpha^2 & \cdots & \alpha^{2(m-1)} \\ \vdots & \vdots & \ddots & \vdots \\ 1 & \alpha^{(\gamma-1)} & \cdots & \alpha^{(\gamma-1)(m-1)} \end{bmatrix}. \quad (8)$$

**Remark 2.** The parity-check matrix $H_{II}$ over $\mathbb{F}_q$ in (7) of $\mathcal{C}_{II}$ is obtained by operations over the same field $\mathbb{F}_q$. In contrast, the parity-check matrix $H$ over $\mathbb{F}_q$ in (4) of a generalized tensor product code is obtained by operations over both the base field $\mathbb{F}_q$ and its extension field.

**Remark 3.** In general, the codes $\mathcal{C}_1$ and $\mathcal{C}_2$ in (6) are chosen to be RS codes [11]. If $\mathcal{C}_1$ and $\mathcal{C}_2$ are chosen to be binary codes, then $m$ can only be $m = 1$.

To see the relation between II codes and generalized tensor product codes, we reformulate $H_{II}$ in (7) into the following form, by *splitting* the rows of $H_2$,

$$H_{II} = \begin{bmatrix} I & \otimes & H_1 \\ \Gamma_2 & \otimes & H_2(1) \\ \Gamma_2 & \otimes & H_2(2) \\ \vdots & \vdots & \vdots \\ \Gamma_2 & \otimes & H_2(k_1 - k_2) \end{bmatrix}, \quad (9)$$

where the matrix $H_1$ over $\mathbb{F}_q$ is the parity-check matrix of $\mathcal{C}_1$, and is treated as a vector over the extension field $\mathbb{F}_{q^{n-k_1}}$ here; correspondingly, the matrix $I$ is treated as an $m \times m$ identity matrix over $\mathbb{F}_{q^{n-k_1}}$. For $1 \leqslant i \leqslant k_1 - k_2$, $H_2(i)$ over $\mathbb{F}_q$ represents the $i$th row of $H_2$, and $\Gamma_2$ over $\mathbb{F}_q$ is the matrix in (8).

Now, referring to the matrix in (4), the matrix in (9) can be interpreted as a parity-check matrix of a $(1 + k_1 - k_2)$-level generalized tensor product code over $\mathbb{F}_q$. Thus, we conclude that an II code is a generalized tensor product code. Using the properties of generalized tensor product codes, we can directly obtain the following result, which was proved in [11] in an alternative way.

**Lemma 10.** *The code $\mathcal{C}_{II}$ is a linear code over $\mathbb{F}_q$ of length $N = nm$, dimension $K = (m - \gamma)k_1 + \gamma k_2$, and minimum distance $D \geqslant \min\{(\gamma+1)d_1, d_2\}$.*

*Proof:* For $1 \leqslant i \leqslant k_1 - k_2$, let the following parity-check matrix
$$\begin{bmatrix} H_1 \\ H_2(1) \\ \vdots \\ H_2(i) \end{bmatrix}$$
define an $[n, k_1 - i, d_{2,i}]_q$ code. It is clear that $d_1 \leqslant d_{2,1} \leqslant d_{2,2} \leqslant \cdots \leqslant d_{2,k_1-k_2} = d_2$.

From the properties of generalized tensor product codes, the redundancy is $N - K = nm - K = (n - k_1)m + \gamma(k_1 - k_2)$; that is, the dimension is $K = k_1(m - \gamma) + k_2 \gamma$. Using Theorem 4, the minimum distance is $D \geqslant \min\left\{d_1(\gamma+1), d_{2,1}(\gamma+1), \ldots, d_{2,k_1-k_2-1}(\gamma+1), d_{2,k_1-k_2}\right\} = \min\left\{(\gamma+1)d_1, d_2\right\}$. ∎

### B. Generalized Integrated Interleaving Codes

With the similar idea used in the previous subsection, we continue our proof for GII codes. We use the definition of GII codes from [24] for consistency. Let $\mathcal{C}_i$, $i = 0, 1, \ldots, \gamma$, be $[n, k_i, d_i]_q$ codes over $\mathbb{F}_q$ such that
$$\begin{aligned} \mathcal{C}_{i_s} = \cdots = \mathcal{C}_{i_{s-1}+1} \subset \mathcal{C}_{i_{s-1}} = \cdots = \mathcal{C}_{i_{s-2}+1} \\ \subset \cdots \subset \mathcal{C}_{i_1} = \cdots = \mathcal{C}_1 \subset \mathcal{C}_0, \end{aligned} \quad (10)$$
where $i_0 = 0$ and $i_s = \gamma$. The minimum distances satisfy $d_0 \leqslant d_1 \leqslant \cdots \leqslant d_\gamma$. A GII code $\mathcal{C}_{GII}$ is defined as:
$$\mathcal{C}_{GII} = \Bigg\{ \mathbf{c} = (\mathbf{c}_0, \mathbf{c}_1, \ldots, \mathbf{c}_{m-1}) : \mathbf{c}_i \in \mathcal{C}_0, 0 \leqslant i < m, \\ \text{and } \sum_{i=0}^{m-1} \alpha^{bi} \mathbf{c}_i \in \mathcal{C}_{\gamma-b}, b = 0, 1, \ldots, \gamma - 1 \Bigg\}, \quad (11)$$
where $\alpha$ is a primitive element of $\mathbb{F}_q$ and $\gamma < m \leqslant q - 1$.

Let us first define some matrices which will be used below. Let the matrix $I$ over $\mathbb{F}_q$ be an $m \times m$ identity matrix. Let $H_0$ over $\mathbb{F}_q$ be the parity-check matrix of $\mathcal{C}_0$. For $1 \leqslant j \leqslant s$, let the matrix $\begin{bmatrix} H_0 \\ H_{i_j} \end{bmatrix}$ over $\mathbb{F}_q$ represent the parity-check matrix of $\mathcal{C}_{i_j}$, where
$$H_{i_j} = \begin{bmatrix} H_{i_1 \setminus i_0} \\ H_{i_2 \setminus i_1} \\ \vdots \\ H_{i_j \setminus i_{j-1}} \end{bmatrix}.$$

For any $i \leqslant j$, let matrix $\Gamma(i, j; \alpha)$ over $\mathbb{F}_q$ be the parity-check matrix of an $[m, m - (j - i + 1), j - i + 2]_q$ code in the following form
$$\Gamma(i, j; \alpha) = \begin{bmatrix} 1 & \alpha^i & \cdots & \alpha^{i(m-1)} \\ 1 & \alpha^{i+1} & \cdots & \alpha^{(i+1)(m-1)} \\ \vdots & \vdots & \ddots & \vdots \\ 1 & \alpha^j & \cdots & \alpha^{j(m-1)} \end{bmatrix}. \quad (12)$$

Now, according to the definition in (11), using the matrices introduced above, the parity-check matrix of $\mathcal{C}_{GII}$ is
$$H_{GII} = \begin{bmatrix} I & \otimes & H_0 \\ \Gamma(0, i_s - i_{s-1} - 1; \alpha) & \otimes & H_{i_s} \\ \Gamma(i_s - i_{s-1}, i_s - i_{s-2} - 1; \alpha) & \otimes & H_{i_{s-1}} \\ \vdots & & \vdots \\ \Gamma(i_s - i_2, i_s - i_1 - 1; \alpha) & \otimes & H_{i_2} \\ \Gamma(i_s - i_1, i_s - i_0 - 1; \alpha) & \otimes & H_{i_1} \end{bmatrix}, \quad (13)$$

which can be transformed into the form of
$$H_{GII} = \begin{bmatrix} I & \otimes & H_0 \\ \Gamma(0, i_s - i_0 - 1; \alpha) & \otimes & H_{i_1 \setminus i_0} \\ \Gamma(0, i_s - i_1 - 1; \alpha) & \otimes & H_{i_2 \setminus i_1} \\ \vdots & & \vdots \\ \Gamma(0, i_s - i_{s-2} - 1; \alpha) & \otimes & H_{i_{s-1} \setminus i_{s-2}} \\ \Gamma(0, i_s - i_{s-1} - 1; \alpha) & \otimes & H_{i_s \setminus i_{s-1}} \end{bmatrix}. \quad (14)$$

To make a connection between GII codes and generalized tensor product codes, we further reformulate the matrix $H_{GII}$ in (14) as follows,

$$H_{GII} = \begin{bmatrix} I & \otimes & H_0 \\ \hline \Gamma(0, i_s - i_0 - 1; \alpha) & \otimes & H_{i_1 \setminus i_0}(1) \\ \vdots & & \vdots \\ \Gamma(0, i_s - i_0 - 1; \alpha) & \otimes & H_{i_1 \setminus i_0}(k_{i_0} - k_{i_1}) \\ \hline \Gamma(0, i_s - i_1 - 1; \alpha) & \otimes & H_{i_2 \setminus i_1}(1) \\ \vdots & & \vdots \\ \Gamma(0, i_s - i_1 - 1; \alpha) & \otimes & H_{i_2 \setminus i_1}(k_{i_1} - k_{i_2}) \\ \hline \vdots & & \vdots \\ \hline \Gamma(0, i_s - i_{s-2} - 1; \alpha) & \otimes & H_{i_{s-1} \setminus i_{s-2}}(1) \\ \vdots & & \vdots \\ \Gamma(0, i_s - i_{s-2} - 1; \alpha) & \otimes & H_{i_{s-1} \setminus i_{s-2}}(k_{i_{s-2}} - k_{i_{s-1}}) \\ \hline \Gamma(0, i_s - i_{s-1} - 1; \alpha) & \otimes & H_{i_s \setminus i_{s-1}}(1) \\ \vdots & & \vdots \\ \Gamma(0, i_s - i_{s-1} - 1; \alpha) & \otimes & H_{i_s \setminus i_{s-1}}(k_{i_{s-1}} - k_{i_s}) \end{bmatrix}, \quad (15)$$

where, in the first level, the matrix $H_0$ over $\mathbb{F}_q$ is treated as a vector over the extension field $\mathbb{F}_{q^{n-k_0}}$, and correspondingly the matrix $I$ is treated as an $m \times m$ identity matrix over $\mathbb{F}_{q^{n-k_0}}$. For $1 \leqslant x \leqslant s$ and $1 \leqslant y \leqslant k_{i_{x-1}} - k_{i_x}$, $H_{i_x \setminus i_{x-1}}(y)$ over $\mathbb{F}_q$ represents the $y$th row of the matrix $H_{i_x \setminus i_{x-1}}$.

Now, referring to the matrix in (4), the matrix in (15) can be seen as a parity-check matrix of a $(1 + k_0 - k_{i_s})$-level generalized tensor product code over $\mathbb{F}_q$. As a result, we can directly obtain the following lemma, which was also proved in [24] in a different way.

**Lemma 11.** *The code $\mathcal{C}_{GII}$ is a linear code over $\mathbb{F}_q$ of length $N = nm$, dimension $K = \sum_{x=1}^{\gamma} k_x + (m - \gamma)k_0 = \sum_{j=1}^{s}(i_j - i_{j-1})k_{i_j} + (m - \gamma)k_0$, and minimum distance $D \geqslant \min\left\{(\gamma+1)d_0, (\gamma - i_1 + 1)d_{i_1}, \ldots, (\gamma - i_{s-1} + 1)d_{i_{s-1}}, d_{i_s}\right\}$.*

*Proof:* For $1 \leqslant x \leqslant s$ and $1 \leqslant y \leqslant k_{i_{x-1}} - k_{i_x}$, let the following parity-check matrix

$$\begin{bmatrix} H_0 \\ \hline H_{i_1 \setminus i_0}(1) \\ \vdots \\ H_{i_1 \setminus i_0}(k_{i_0} - k_{i_1}) \\ \hline \vdots \\ \hline H_{i_x \setminus i_{x-1}}(1) \\ \vdots \\ H_{i_x \setminus i_{x-1}}(y) \end{bmatrix}$$

define an $[n, k_{i_{x-1}} - y, d_{i_x, y}]_q$ code, so we have $d_{i_{x-1}} \leqslant d_{i_x,1} \leqslant d_{i_x,2} \leqslant \cdots \leqslant d_{i_x, k_{i_{x-1}} - k_{i_x}} = d_{i_x}$. From the properties of generalized tensor product codes, it is easy to obtain the dimension $K = \sum_{j=1}^{s}(i_j - i_{j-1})k_{i_j} + (m - \gamma)k_0$. From Theorem 4, the minimum distance satisfies

$$\begin{aligned} D \geqslant \min \Big\{ & (\gamma + 1)d_0, \\ & (\gamma + 1)d_{i_1, 1}, \ldots, (\gamma + 1)d_{i_1, k_{i_0} - k_{i_1} - 1}, (\gamma - i_1 + 1)d_{i_1}, \\ & \ldots, \ldots, (\gamma - i_{s-1} + 1)d_{i_{s-1}}, \\ & (\gamma - i_{s-1} + 1)d_{i_s, 1}, \ldots, (\gamma - i_{s-1} + 1)d_{i_s, k_{i_{s-1}} - k_{i_s} - 1}, d_{i_s} \Big\} \\ = \min \Big\{ & (\gamma + 1)d_0, (\gamma - i_1 + 1)d_{i_1}, \\ & \ldots, (\gamma - i_{s-1} + 1)d_{i_{s-1}}, d_{i_s} \Big\}. \end{aligned}$$

∎

**Remark 4.** In some prior works, we find that generalized tensor product codes are called generalized error-location (GEL) codes [4], [16]. Recently, in [24], the similarity between GII codes and GEL codes was observed. However, the exact relation between them was not studied. In [24], the author also proposed a new generalized integrated interleaving scheme over binary BCH codes, called GII-BCH codes. These codes can also be seen as a special case of generalized tensor product codes.

## VI. CONCLUSION

In this work, we presented a general construction for ME-LRCs over small fields. This construction yields optimal ME-LRCs with respect to an upper bound on the minimum distance for a wide range of code parameters. Then, an erasure decoder was proposed and corresponding correctable erasure patterns were identified. ME-LRCs based on Reed-Solomon codes were shown to be optimal among all codes having the same erasure-correcting capability. Finally, generalized integrated interleaving codes were proved to be a subclass of generalized tensor product codes, thus giving the exact relation between these two codes.

## ACKNOWLEDGMENT

This work was supported by NSF Grants CCF-1405119 and CCF-1619053, BSF Grant 2015816, and Western Digital Corporation.

## APPENDIX A
## PROOF OF LEMMA 2

*Proof:* For the case of $x = 0$, it is trivial. For $1 \leqslant x \leqslant \lceil \frac{k}{k^*} \rceil - 1$, $x \in \mathbb{Z}^+$, let $\mathcal{I}$ represent the set of the coordinates of the first $x$ rows in the array. Thus, $|\mathcal{I}| = xn_0$. First, consider the code $\mathcal{C}_\mathcal{I} = \{c_\mathcal{I} : c \in \mathcal{C}\}$ whose dimension is denoted by $k_\mathcal{I}$, which satisfies $k_\mathcal{I} \leqslant xk^*$. Then, we consider the code $\mathcal{C}_\mathcal{I}^0 = \{c_{[\rho n_0] \setminus \mathcal{I}} : c_\mathcal{I} = 0 \text{ and } c \in \mathcal{C}\}$. Since the code $\mathcal{C}$ is linear, the size of the code $\mathcal{C}_\mathcal{I}^0$ is $q^{k-k_\mathcal{I}}$ and it is a linear code as well. Moreover, the minimum distance $\hat{d}$ of the code $\mathcal{C}_\mathcal{I}^0$ is at least $d$, i.e., $\hat{d} \geqslant d$.

Thus, we get an upper bound on the minimum distance $d$,

$$d \leqslant \hat{d} \leqslant d_{opt}^{(q)}[\rho n_0 - |\mathcal{I}|, k - k_\mathcal{I}]$$
$$\leqslant d_{opt}^{(q)}[\rho n_0 - xn_0, k - xk^*].$$

Similarly, we also get an upper bound on the dimension $k$,

$$k - k_\mathcal{I} \leqslant k_{opt}^{(q)}[\rho n_0 - |\mathcal{I}|, \hat{d}] \leqslant k_{opt}^{(q)}[\rho n_0 - xn_0, d].$$

Therefore, we conclude that

$$k \leqslant k_{opt}^{(q)}[\rho n_0 - xn_0, d] + k_\mathcal{I} \leqslant k_{opt}^{(q)}[\rho n_0 - xn_0, d] + xk^*.$$

∎

## APPENDIX B
## PROOF OF LEMMA 3

*Proof:* We can construct a $(\rho, n_0, k; \geqslant d_0, \geqslant d)_q$ ME-LRC in two steps, and use the GV bound [20] twice. First, there exists a $[\rho(n_0 - r_0), k, \geqslant d]_q$ array code $\mathcal{G}_1$ of size $\rho \times (n_0 - r_0)$ where $r_0$ is an integer $0 \leqslant r_0 < n_0$, if it satisfies

$$\sum_{i=0}^{d-2} \binom{\rho(n_0 - r_0) - 1}{i}(q-1)^i < q^{\rho(n_0 - r_0) - k}. \quad (16)$$

Second, there exists a length-$n_0$ code $\mathcal{G}_2$ with minimum distance at least $d_0$, if its redundancy $r_0$ satisfies

$$r_0 > \log_q \left( \sum_{i=0}^{d_0-2} \binom{n_0 - 1}{i}(q-1)^i \right). \quad (17)$$

Now, we encode each row of the code $\mathcal{G}_1$ using the code $\mathcal{G}_2$ by adding $r_0$ more redundancy symbols. The resulting code is a $(\rho, n_0, k; \geqslant d_0, \geqslant d)_q$ ME-LRC. Let $r_0 = \lceil \log_q \left( \sum_{i=0}^{d_0-2} \binom{n_0-1}{i}(q-1)^i \right) \rceil$, and substitute it into (16), producing (3). ∎

## APPENDIX C
## PROOF OF THEOREM 4

*Proof:* A codeword $x$ in $\mathcal{C}_{GTP}^\mu$ is an $n'\ell$-dimensional vector over $\mathbb{F}_q$, denoted by $x = (x_1, x_2, \ldots, x_\ell)$, where $x_i$ in $x$ is an $n'$-dimensional vector, for $i = 1, 2, \ldots, \ell$.

Let $s_i^j = x_i {H'_j}^T$, for $i = 1, 2, \ldots, \ell$ and $j = 1, 2, \ldots, \mu$. Thus, $s_i^j$ is a $v_j$-dimensional vector over $\mathbb{F}_q$, and is considered as an element in $\mathbb{F}_{q^{v_j}}$. Let $s^j = (s_1^j, s_2^j, \ldots, s_\ell^j)$, an $\ell$-dimensional vector over $\mathbb{F}_{q^{v_j}}$, whose components are $s_i^j$, $i = 1, 2, \ldots, \ell$.

To prove Theorem 4, we need to show that if $xH^T = 0$ and $w_q(x) < d_m = \min\{\delta_1, \delta_2 d'_1, \delta_3 d'_2, \ldots, \delta_\mu d'_{\mu-1}, d'_\mu\}$, then $x$ must be the all-zero vector $\mathbf{0}$.

We prove it by contradiction and induction. Assume that there exists a codeword $x$ such that $xH^T = 0$, $w_q(x) < d_m$, and $x \neq \mathbf{0}$.

We first state a proposition which will be used in the following proof.

**Proposition 12.** *If $xH^T = 0$ and $s^1 = s^2 = \cdots = s^j = \mathbf{0}$, then $w_q(x_i) \geqslant d'_j$ for $x_i \neq \mathbf{0}$, $i = 1, 2, \ldots, \ell$.*

*Proof:* The condition $s^1 = s^2 = \cdots = s^j = \mathbf{0}$ means that $x_i B_j^T = \mathbf{0}$ for $i = 1, 2, \ldots, \ell$; that is, $x_i$ is a codeword in the code defined by the parity-check matrix $B_j$, whose minimum distance is $d'_j$. Therefore, we have $w_q(x_i) \geqslant d'_j$ for $x_i \neq \mathbf{0}$, $i = 1, 2, \ldots, \ell$. ∎

Now, if $s^1 \neq \mathbf{0}$, then $w_q(x) \geqslant w_{q^{v_1}}(s^1) \geqslant \delta_1 \geqslant d_m$, which contradicts the assumption. Thus, we have $s^1 = \mathbf{0}$.

Then, consider the second level. If $s^2 \neq \mathbf{0}$, then $w_q(x) \overset{(a)}{\geqslant} w_{q^{v_2}}(s^2) d'_1 \geqslant \delta_2 d'_1 \geqslant d_m$, where step (a) is from Proposition 12. This contradicts the assumption, so we have $s^2 = \mathbf{0}$. By induction, we must have $s^1 = s^2 = \cdots = s^{\mu-1} = \mathbf{0}$.

For the last level, i.e., the $\mu$th level, if $s^\mu \neq \mathbf{0}$, then $w_q(x) \geqslant w_{q^{v_\mu}}(s^\mu) d'_{\mu-1} \geqslant \delta_\mu d'_{\mu-1} \geqslant d_m$, which contradicts our assumption. Now, if $s^1 = s^2 = \cdots = s^\mu = \mathbf{0}$, then $w_q(x) \geqslant d'_\mu \geqslant d_m$, which also contradicts our assumption.

Thus, our assumption is violated. ∎

## APPENDIX D
## PROOF OF THEOREM 6

*Proof:* The proof follows from the decoding procedure of decoder $\mathcal{D}_A$. The ME-LRC $\mathcal{C}_A$ has $d_0 = d'_1$ and $d = d'_\mu$. For a received word $y = (y_1, y_2, \ldots, y_\ell)$, each vector $y_i$, $1 \leqslant i \leqslant \ell$, corresponds to a row in the array.

For the first level, since $\delta_1 = \infty$, the correct syndrome vector $(s_1^1, \ldots, s_\ell^1)$ is the all-zero vector, i.e., $(s_1^1, \ldots, s_\ell^1) = \mathbf{0}$. Thus, the rows with number of erasures less than $d'_1$ are corrected.

For the second level, the remaining uncorrected row $\hat{c}_j$, $j \in \mathcal{F}$, has at least $d'_1$ erasures. The total number of such uncorrected rows with indices in $\mathcal{F}$ is less than $\delta_2$, because we require $N_1 \leqslant \delta_2 - 1$ in the condition. Thus, the correct syndrome vector $(s_1^2, \ldots, s_\ell^2)$ can be obtained. As a result, the rows with number of erasures less than $d'_2$ are corrected.

Similarly, by induction, if the decoder runs until the $\mu$th level, the remaining uncorrected row $\hat{c}_j$, $j \in \mathcal{F}$, has at least $d'_{\mu-1}$ erasures. The total number of such uncorrected rows with indices in $\mathcal{F}$ is less than $\delta_\mu$, because we require $N_{\mu-1} \leqslant \delta_\mu - 1$ in the condition. Therefore, all the correct syndrome vectors $(s_1^i, \ldots, s_\ell^i)$, $i = 1, 2, \ldots, \mu$, are obtained. On the other hand, the remaining uncorrected row $\hat{c}_j$, $j \in \mathcal{F}$, has at most $d'_\mu - 1$ erasures, since we also require $N_\mu \leqslant 0$ in the condition. Thus, all these uncorrected rows can be corrected in this step with all these correct syndromes. ∎

## APPENDIX E
## PROOF OF COROLLARY 7

*Proof:* The ME-LRC $\mathcal{C}_A$ has $d_0 = d'_1$ and $d = d'_\mu$. We only need to show that the received word $y$ with any $d'_\mu - 1$ erasures satisfies the condition in Theorem 6. We prove it by contradiction. If the condition is not satisfied, there is at least an integer $i$, $1 \leqslant i \leqslant \mu$, such that $N_i \geqslant \delta_{i+1}$. Therefore, we have $w_e(y) \geqslant d'_i \delta_{i+1} \geqslant d'_\mu$, where the last inequality is from the requirement of Construction A. Thus, we get a contradiction to the assumption that the received word $y$ has $d'_\mu - 1$ erasures. ∎